\documentclass[conference]{IEEEtran}

\makeatletter
\def\ps@IEEEtitlepagestyle{%
  \def\@oddfoot{\mycopyrightnotice}%
  \def\@evenfoot{}%
}
\def\mycopyrightnotice{%
  {\footnotesize 978-1-6654-7095-7/22/\$31.00~\copyright~2022 IEEE\hfill}
  \gdef\mycopyrightnotice{}
}

\usepackage{algorithm2e}
\usepackage{tikz}
\usepackage{blindtext}
\usepackage{eso-pic}
\IEEEoverridecommandlockouts
\usepackage{cite}
\usepackage{amsmath,amssymb,amsfonts}
\usepackage{algorithmic}
\usepackage{graphicx}
\usepackage{textcomp}
\usepackage{url}
\usepackage{xcolor}
\def\BibTeX{{\rm B\kern-.05em{\sc i\kern-.025em b}\kern-.08em
    T\kern-.1667em\lower.7ex\hbox{E}\kern-.125emX}}
    
\usepackage{eso-pic}
\newcommand\AtPageUpperMyright[1]{\AtPageUpperLeft{%
 \put(\LenToUnit{0.17\paperwidth},\LenToUnit{-2cm}){%
     \parbox{0.9\textwidth}{\raggedleft\fontsize{8}{11}\selectfont #1}}%
 }}%
\newcommand{\conf}[1]{%
\AddToShipoutPictureBG*{%
\AtPageUpperMyright{#1}
}
} 

\begin{document}
\title{\vspace*{1cm} Blockchain-based Monitoring for Poison Attack Detection in Decentralized Federated Learning
}

\author{\IEEEauthorblockN{Ranwa Al Mallah}
\IEEEauthorblockA{\textit{Electrical and Computer Engineering} \\
\textit{Royal Military College of Canada}\\
Kingston, Canada \\
ranwa.al-mallah@rmc-cmr.ca}
\and
\IEEEauthorblockN{David López}
\IEEEauthorblockA{\textit{Instituto de Ingeniería} \\
\textit{Universidad Nacional Autónoma de México}\\
Mexico City, Mexico \\
dlopezfl@iingen.unam.mx}
}

\maketitle
\conf{\textit{  Proc. of the International Conference on Electrical, Computer, Communications and Mechatronics Engineering  (ICECCME) \\ 
16-18 November 2022, Maldives}}
\begin{abstract}
Federated Learning (FL) is a machine learning technique that addresses the privacy challenges in terms of access rights of local datasets by enabling the training of a model across nodes holding their data samples locally. To achieve decentralized federated learning, blockchain-based FL was proposed as a distributed FL architecture. In decentralized FL, the \textit{chief} is eliminated from the learning process as \textit{workers} collaborate between each other to train the global model. Decentralized FL applications need to account for the additional delay incurred by blockchain-based FL deployments. Particularly in this setting, to detect targeted/untargeted poisoning attacks, we investigate the end-to-end learning completion latency of a realistic decentralized FL process protected against poisoning attacks. We propose a technique which consists in decoupling the monitoring phase from the detection phase in defenses against poisoning attacks in a decentralized federated learning deployment that aim at monitoring the behavior of the \textit{workers}. We demonstrate that our proposed blockchain-based monitoring improved network scalability, robustness and time efficiency. The parallelization of operations results in minimized latency over the end-to-end communication, computation, and consensus delays incurred during the FL and blockchain operations. 
\end{abstract}

\section{Introduction}
\label{introduction}

In traditional centralized machine learning, the nodes participating in the training of a model upload their local datasets to a central node \cite{verbraeken2020survey}. Distributed machine learning is a system to train independently a model on different nodes. This distributed system accelerates the training for very big amounts of data. However, since data might be sensitive, we need a privacy preserving distributed learning system for collaborative training of distributed machine learning models without any data sharing. Federated learning is a machine learning paradigm that addresses the privacy, security and access rights challenges related to local datasets by training a global model across decentralized nodes and enabling the \textit{worker} nodes to hold their data samples locally without sharing them \cite{yang2019federated}.

In centralized federated learning, a central node, the \textit{chief}, manages the different steps of the learning process that the \textit{workers} have to go through to train the global model. The \textit{chief} is mainly responsible for the coordination of the \textit{workers} and aggregation of the received local model updates. However, in some settings, this strategy may constitute a bottleneck since all the \textit{workers} have to send their local model updates to the \textit{chief}, thus affecting the performance of the \textit{chief}.

 In decentralized federated learning, the \textit{workers} collaborate between each other to train the global model. This strategy eliminates the \textit{chief} from the learning process and prevents the single point of failure that the \textit{chief} represents in centralized federated learning. 

 To achieve decentralized federated learning, researchers have developed blockchain-based federated learning as a distributed FL architecture because both the blockchain and federated learning technologies protect the privacy of individuals. The blockchain is a crypto-based secure ledger for data storage and transfer through decentralized, trustless peer-to-peer systems. In our setting, the blockchain network will enable the exchange of the \textit{workers}’ local model updates while a subset of \textit{workers}, the miners in the context of a blockchain, are responsible of the aggregation of the received local model updates. The miners can be either randomly selected nodes or separate nodes as in a traditional blockchain network. The \textit{workers} will send their local model updates to their associated miner in the blockchain network. The miners will exchange between each other all the local model updates in order to proceed with the training of the model. The miners can then complete the first round of the process by running the consensus algorithm. Consensus results in a block to be added to the blockchain and the block stores the global model which is the aggregate of the received local model updates. In the second round of the process, the  \textit{Workers} will then download the global model from the blockchain. The global model serves as an input to the next local model update that the \textit{worker} will generate in the next iteration of the federated learning. 

 Decentralized federated learning applications need to account for the additional delay incurred by the blockchain network. Many studies aim at addressing the latency challenges related to decentralized federated learning applications \cite{kim2018device}. One application that must consider the end-to-end latency incurred by blockchain-based decentralized federated learning is the detection of targeted/untargeted poisoning attacks. To the best of our knowledge, no previous work has addressed this challenge in this setting. To this aim, we propose a technique to address the latency challenges and the technique consists in the decoupling of the monitoring phase from the detection phase in decentralized FL approaches defenses that protect against poisoning attacks in federated learning. 

 Poisoning attacks in federated learning happen when the attacker is able to inject malicious data into the model during training, and hence alter the learning process. Steinhardt et al. \cite{steinhardt2017certified} reported that, even under strong defenses, a $3\%$ training dataset poisoning leads to $11\%$ drop in accuracy. Regarding the cybersecurity of FL and protection against this type of attack, many defenses were proposed for the centralized federated learning setting \cite{abadi2016tensorflow}, \cite{blanchard2017machine}, \cite{chen2017distributed}, \cite{mhamdi2018hidden}. Lately, detection and behavioral pattern analysis as a defense mechanisms are gaining momentum. The aim of this type of defense is to remove at every iteration of the FL process, the unreliable nodes in the system based on the assessment of their behavior \cite{pan2020justinian}. By monitoring the behavior in time of the \textit{workers} and removing unreliable nodes from the aggregation process, the approaches enable efficiency and security of the centralized FL process. 

 No studies have yet addressed the additional delay incurred by the detection of targeted/untargeted poisoning attacks techniques in a decentralized federated learning setting. In fact, network topologies affect the performances of the learning process in distributed FL architectures. We propose a technique where monitoring and detection is done in parallel by the blockchain that returns a filtered reliable set of \textit{workers} from which miners in the decentralized federated learning environment can randomly pick a subset to continue the FL process. We improved security, scalabilty and time efficiency because the monitoring is decoupled from the FL and distributed. The technique shows great levels of time efficiency because it leverages, on one hand, the underlying blockchain as an immutable security monitoring system for the \textit{workers}, and on the other hand the latency is minimized in the detection phase because verification is done separately of the decentralized federated learning process. 

 The rest of the paper is organised as follows. In Section~\ref{sec:related_work} we present related work. In Section~\ref{sec:blockchain-based_defense} we present the blockchain-based monitoring. We provide experimental results of the poisoning attacks and defense on a mode inference model implemented over the blockchain as a case study in Section~\ref{sec:evaluation}. We conclude the paper and provide future work in Section~\ref{sec:conclussions}.

\section{Related Work}
\label{sec:related_work}

In federated learning, a star network where a central server (\textit{chief}) is connected to a network of nodes, is the predominant communication topology. However, decentralized topologies (where nodes communicate with their neighbors) are a potential alternative. In data center environments and under a particular setting, decentralized training has been demonstrated to be faster than centralized training when operating on networks with low bandwidth or high latency. Similarly, in federated learning, decentralized algorithms can in theory reduce the high communication cost on the central server. Hierarchical communication patterns have also been proposed \cite{liu2019edge} to further ease the burden on the \textit{chief}, by first leveraging edge servers to aggregate the updates from edge nodes and then relying on a cloud server to aggregate updates from edge servers. While this is a promising approach to reduce communication, it is not applicable to all networks, as this type of physical hierarchy may not exist or be known a priori.

A blockchain can be used to replace the \textit{chief}, the centralized aggregator in the traditional FL system. Miners calculate the averaged model using received update models from \textit{workers}. 

 As an example application, L\'opez et al. \cite{lopez2018blockchain} presented a Blockchain for Smart Mobility Data-markets for mobility data transactions designed to solve the privacy, security, and management issues related to the sharing of passively or actively solicited large-scale data. They developed a federated learning environment over that blockchain to create a privacy-aware solution for mode inference \cite{Lopez2019BCFL}. They show that nodes that collectively, but privately train a Convolutional Neural Network (CNN) can achieve the same accuracy as the conventional centralized training.

 Preuveneers et al. \cite{preuveneers2018chained} proposed a decentralized federated learning environment where model updates are stored on the distributed ledger. This approach exudes large computation overload and latencies. Also, their solution only guarantee the immutability and verify the integrity of the stored data but cannot assure its veracity. There is no defense mechanism implemented as a security measure to protect against poisoning attacks in this setting. Unlike their approach, our technique is applied on decentralized federated learning processes where the defense is integrated in the training of the algorithm in order to eliminate malicious \textit{workers} that compromise the learning. Qu et al. \cite{qu2020decentralized} used the blockchain to enable FL without any centralized authority but they provided incentives to the \textit{workers}, which may result in biased training of the models as only a particular type of \textit{worker} might be interested in the process. In their work, to guarantee block generation efficiency, pointers of the averaged model are saved on-chain while a distributed hash table is used to save the data and point to an off-chain data storage. Again, they do not consider a defense mechanism implemented to secure the undergoing FL process.

 Zhao et al. \cite{zhao2019mobile} introduced the concept of reputation and reputation status is recorded in the blockchain. However, to verify the validity of a model update, they only check if the signature is invalid, the miner rejects the transaction. They do not study the quality of the uploaded model updates and do not defend against poisoning attacks. Kang et al. \cite{kang2020reliable} also use the reputation to reflect how well a \textit{worker} has performed about model training, which is measured from its training task completion history with the past behaviors of good or unreliable activities. To remove malicious updates, they use the FoolsGold scheme \cite{fung2018mitigating}. FoolsGold is a defense against targeted poisoning attacks based on inter-client contribution similarity in their model updates. 

 Some defense mechanisms consider the behavior in time of the \textit{workers} during training in order to detect an anomaly and reject poisonous model updates \cite{pan2020justinian}. With a temporal and dynamic monitoring method, the \textit{chief} can detect and remove malicious or unreliable \textit{workers} from the system. In this context, the blockchain is used to monitor the training of individual \textit{workers} and at the same time, detect a malicious \textit{worker}, thus, ensuring that transactions stored on the blockchain are valid. The inclusion of the blockchain in this setting aims at improving robustness of the system against the attacks. However, this will affect the performance of the learning process in terms of latency. Some studies propose a latency analysis of federated learning via blockchain \cite{kim2018device, mcmahan2017communication, samarakoon2019distributed}. However, the approaches are not applicable in the context of a cybersecurity layer implemented on top of the decentralized federated learning process occurring between the nodes. When a detection mechanism is integrated to protect against poisoning attacks in a decentralized FL process, and particularly when the defense requires monitoring of the behavior of the \textit{workers}, to account for the additional latency generated, we propose a technique where monitoring is decoupled from detection.

\section{Blockchain-based Monitoring}
\label{sec:blockchain-based_defense}

\begin{figure}[h]
\begin{center}
  \includegraphics[width=\columnwidth]{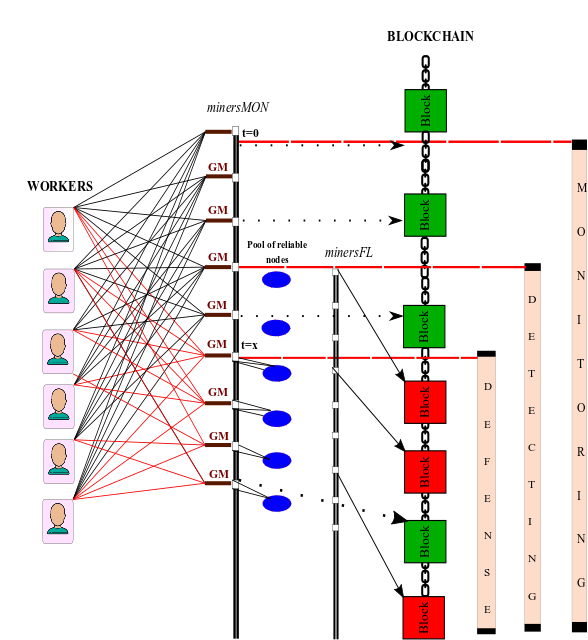}
  \caption{Schema of the blockchain-based monitoring, detection and training process.}
  \label{fig:Parallelprocess}
\end{center}
\end{figure}

We use the blockchain to develop an immutable infrastructure for the decentralized federated learning process of a model and for the monitoring of the \textit{workers} to detect poisoning attacks. In order to train the shared model efficiently, the security layer added to the blockchain should not constitute a bottleneck by affecting the time taken for model training. In fact, we show in Figure~\ref{fig:Parallelprocess} the design we put forward that results in high levels of time efficiency. Some miners of the blockchain network are attributed to the FL process and they are called, \textit{minersFL} and others, \textit{minersMON}, are attributed to the monitoring. When training starts at \textit{t}=0, \textit{workers} in the distributed environment perform the federated learning process as usual. They extract from the green block that was already in the blockchain, the randomly initialized model that they will use to train with their local data. In the initial stage, before detection and defense kicks in, all \textit{workers} continuously send, iteration after iteration, their local model updates to the subset of miners called \textit{minersMON}. It’s only after some period of time \textit{t < x} that the blockchain-based monitoring algorithm is able to return a filtered reliable set of \textit{workers} from which \textit{minersFL} nodes can randomly pick a subset to continue the FL process. It is at time \textit{t = x} precisely that the system becomes protected against attackers although since the beginning, monitoring was performed. Afterwards, iteration after iteration, only reliable nodes will be selected from the pool of nodes to participate in the training. Thus, monitoring and detection are done separately from training, but they must be in synchronisation with each other so as to always provide fresh and accurate evaluation of the nodes. 

 It is important to note that in our design, the model updates are not written on the blockchain, instead, a hash value inside the public ledger points towards them. The model updates remain on the \textit{worker} side encrypted with keys that no other nodes have access to. Digital signatures are required for information to be stored on the blockchain. Therefore, we hold that attackers are not able to fabricate digital signatures or take control of the majority of the network (over 50\%) in order to modify valid blocks on the blockchain. Furthermore, an attacker cannot poison the training data samples because they are stored off-chain on the nodes rather than on the public ledger. However, if the attacker gains access and controls one or some of the nodes, it can send model updates on their behalf that are maliciously fabricated to compromise the system. We detail the blockchain-based monitoring in Algorithm~\ref{alg:BB}.

\begin{algorithm}[h]
\SetAlgoLined
\KwResult{$H_{i,z}$ a subset \textit{z} of the \textit{worker} \textit{i} previously uploaded consecutive Local Model update recorded as a pair of $LM$ and $GM$ at that time\;}
 {\bfseries Input:} Global Model $GM^t$ at iteration \textit{t}\; 
   \For{Iteration \textit{t}}{
   \textit{minersFL} select a subset of \textit{workers} and sends them $GM^t$\; 
   \textit{workers} train, encrypt their $LM^{t + 1}$, perform hash and insert it in Merkle tree on the node\;
   \textit{workers} sign Merkle root and transmit it to the \textit{minersMON} nodes of the blockchain\;
   \textit{minersMON} verify identities before storing signed Merkle root on the blockchain\; 
   \For{$\forall$ \textit{workers} \textit{i}}{
   \textit{minersMON} extract Merkle root and select a random time window to examine the behavior of the \textit{workers}\;
   \textit{workers} send to the \textit{minersMON} the required model updates and their corresponding Merkle path as a proof for examination\;
    \If{Hashes are valid}{
    \textit{minersMON} compute $H_{i,z}$\;
   \textit{minersMON} return to \textit{minersFL} a filtered reliable set of \textit{workers}\;
   \textit{minersFL} randomly pick a subset of nodes from the filtered reliable set to perform the aggregation step of the FL process\;
    }
   }
   }
\caption{Blockchain-based monitoring}
 \label{alg:BB}
\end{algorithm}

 The aim is to store on the blockchain, a commitment on behalf of the \textit{worker} on a series of model updates it had worked on. The order in time of the model updates is important. We will use a Merkle Tree because the detection algorithm needs to go back to previous model updates of a \textit{worker} to validate a benign behavior \cite{szydlo2004merkle}. The algorithm might want to selectively look at a portion in time of the submitted model updates without requiring the \textit{worker} to send all the models updates it had since the beginning of training. At the same time, this forces the \textit{worker} to commit to all the model updates it ever did without knowing which portion is going to be evaluated by the miners of the blockchain network. Most importantly, the \textit{worker} will never be able to go back and modify an entry because the linked timestampting of the hash chain in the blockchain is a commitment to every previous value. Also, the \textit{worker} will not be able to reorder previous entries because of the binding commitment property to all the messages. Using a Merkle Tree is efficient both because it enables to store on the blockchain, a Merkle root of only 256 bits and enables to selectively reveal a portion of model updates. The Merkle root is a commitment to the Merkle Tree of the entire set of a node's model updates. For the miners to open the commitment to a single model update, all it requires is its Merkle path. 

\begin{figure}[h]
\begin{center}
  \includegraphics[width=\columnwidth]{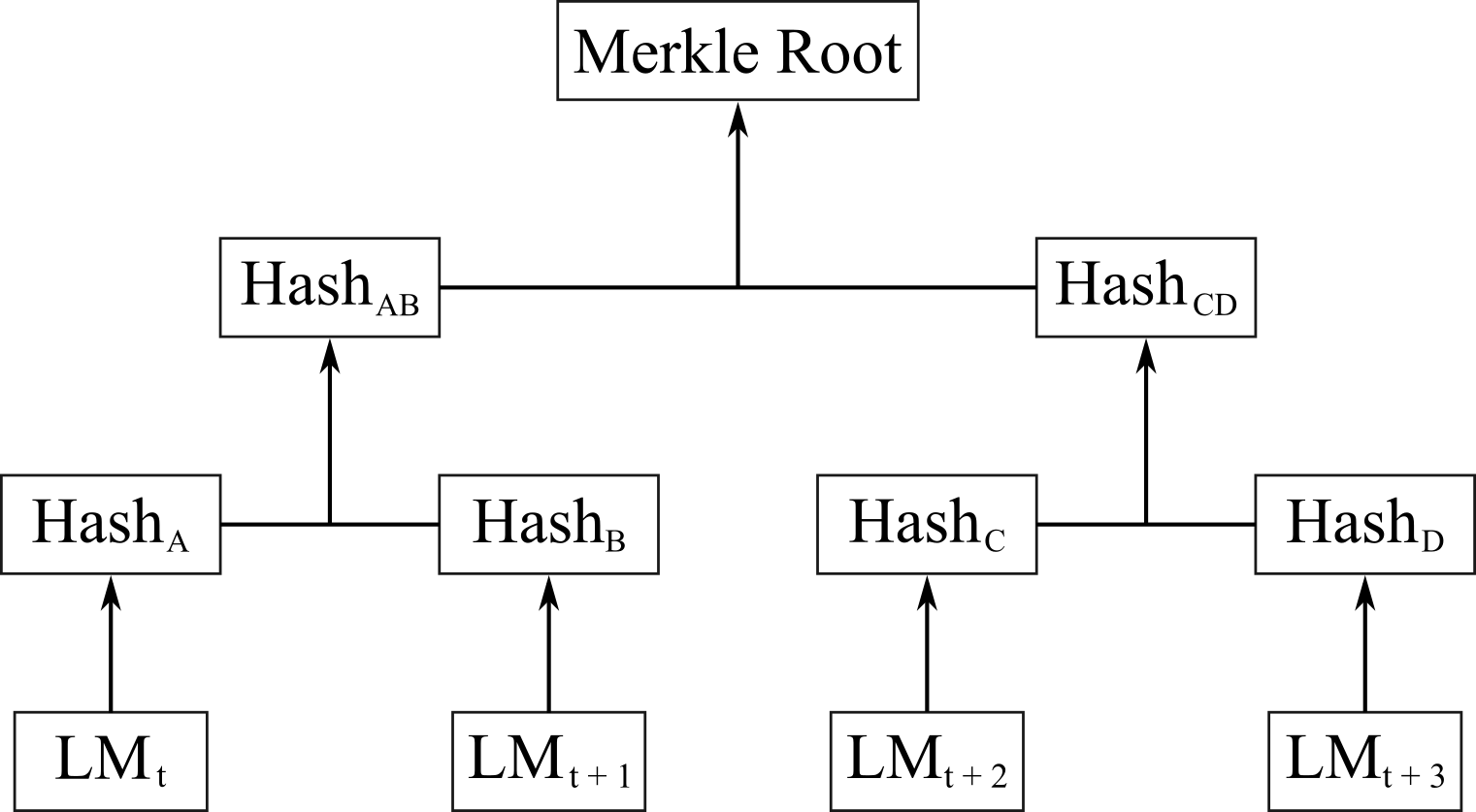}
  \caption{Local Model updates stored in a Merkle tree.}
  \label{fig:MerkleDesign}
\end{center}
\end{figure}

 We investigated different designs. For example, a hash function, SHA256 takes a variable input size, but always returns a 256-bit hash value. The hash of one model update is 256 bits. However, at each round, storing on the blockchain the hash of every previous model update is linear in input size. Another design would be to store the hash of the concatenation of all previous model updates. This would be constant in size no matter the input. However, in this case, in order to verify the validity of one model update, all other concatenated model updates must be provided. For this reason, using a Merkle Tree is more efficient than the concatenation because it enables to store on the blockchain, a Merkle root of 256 bits and enables to selectively reveal a portion of model updates. We show in Figure~\ref{fig:MerkleDesign} a prototype of the Merkle Tree stored at the \textit{worker} node with the Merkle root being the binding commitment to the entire model updates of the node during training. Merkle trees allow efficient and secure verification of the contents of large data structures. This design achieves scalability and most importantly, enables storage of valuable data in contrast to approaches that store reputation metrics. Our design stores the underling temporal and dynamic local models updates of every \textit{worker} of the system in a transparent and secure way. The monitoring and the training are done in parallel by different miners of the blockchain network.

\section{Evaluation}
\label{sec:evaluation}
To evaluate the efficiency of our technique in addressing the latency challenges, we first implement a blockchain-based decentralized FL process consisting in training a Convolutional Neural Network classifier for transportation mode inference as in \cite{Lopez2019BCFL}. We then add the security layer for protection against poisoning attacks under this setting. Any defense that aim at monitoring the behavior of the \textit{workers} can be implemented. We perform a poisoning attack on the system and compute the time taken by the FL process to converge. We then implement our technique consisting in the decoupling of the monitoring and detection phases. We compare the latency incurred over the end-to-end communication, computation, and consensus delays incurred during the FL and blockchain operations. We demonstrate that the parallelization of operations results in minimized latency. 

\subsection{Case study}
The dataset of this study consists of raw personal data of nodes participating in the task of distributed behavioural choice modelling over the blockchain. As in \cite{Lopez2019BCFL}, each observation is characterized with trip duration, trip reliability and trip cost. For testing the distributed choice model, we use a subset of 246 observations, only including automobile and train as the two mode choices to explore how choice modelling can be distributed over a blockchain. In this context, the nodes are always in control of their data because they don't share their raw personal information with anyone. Although this approach ensures the protection of the individual's privacy, it exposes the system to poisoning attacks. In fact, malicious nodes can inject maliciously fabricated local model updates to sabotage the choice modeling training that is ongoing. This will poison the learning process.

\subsection{Experimental setup}
The experiments are implemented on four nodes that run on \emph{Hyperledger Iroha}. The nodes are responsible of storing the blockchain and some of them are in charge of different operations as miners participating in different consensus mechanisms. Each node is running on an \emph{Amazon EC2 t2.medium Virtual Machine}. One \emph{Amazon EC2 t3.2xlarge Virtual Machine} is used to run the 10 nodes who are participating in the FL process. Some nodes are designated as \textit{minersFL} and others as \textit{minersMON}, while all can also behave as \textit{workers} in the decentralized setting.

\subsection{Experimental Results}

\begin{figure}[h]
\begin{center}
  \includegraphics[width=\columnwidth]{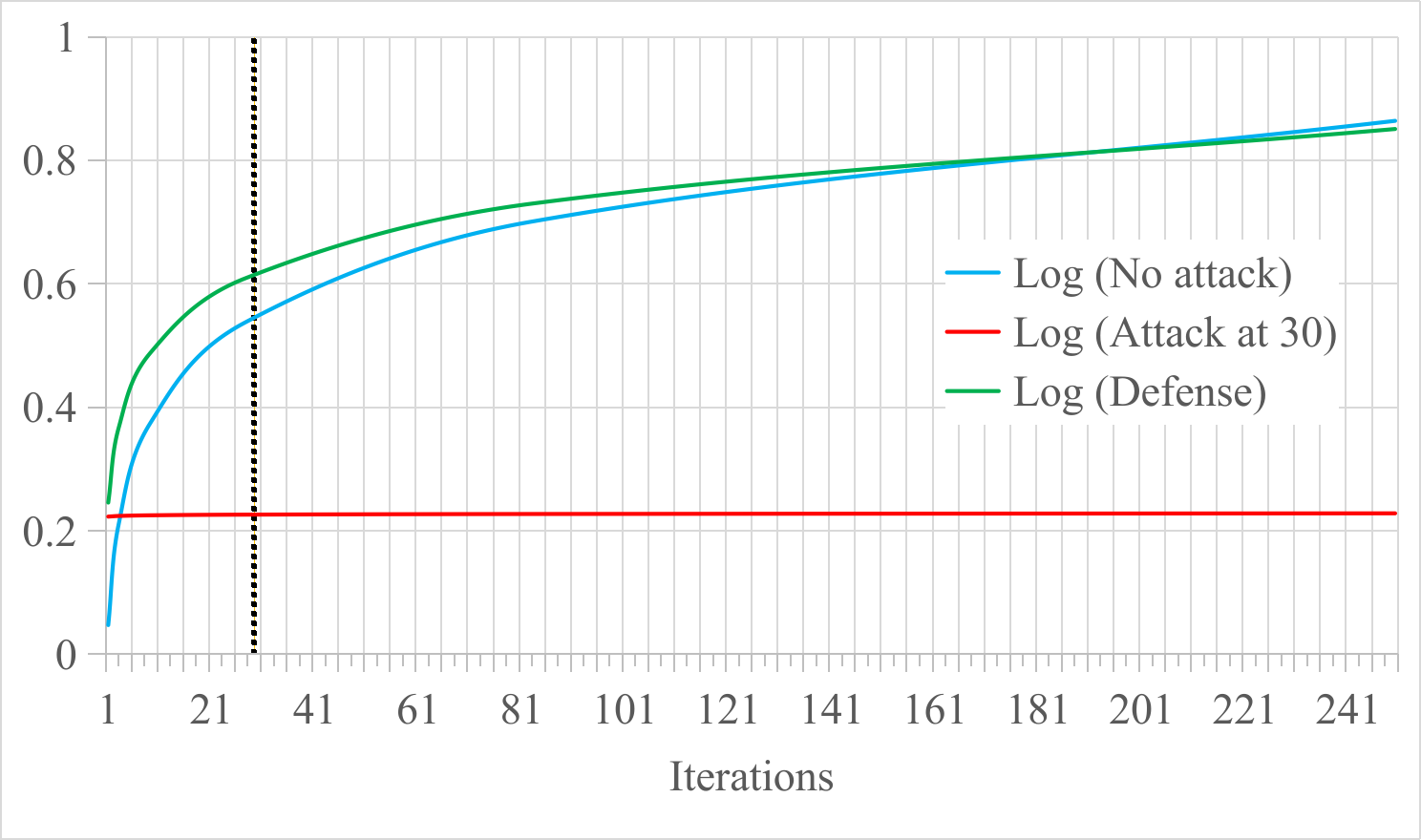}
  \caption{Latency in terms of convergence time of a decentralized FL process for the training of a classifier for transportation mode inference: under no attack, under attack: one attacker, parallelization defense technique.}
  \label{fig:Time_FL_one}
\end{center}
\end{figure}

We show in Figure~\ref{fig:Time_FL_one} the convergence time taken by the blockchain-based decentralized FL process for the training of a classifier for transportation mode inference under normal conditions (under no attack). We then implement a poisoning attack where one attacker performs a continuous untargeted attack after EPOCH $30$ by injecting random weight updates aiming at decreasing convergence speed and compromising the system. We compare in the same figure the time taken by the FL process to converge under attack. We notice that as the number of iterations was increasing under attack, the system was never able to converge. 

We then implement our technique consisting in the decoupling of the monitoring and detection phases as a defense aiming at monitoring the behavior of the \textit{workers} in the system as per Algorithm 1. In the same figure, we present the results in terms of latency incurred. Compared to the scenario under attack, we notice that the system was able to converge and we see how the end-to-end delay decreased. The parallelization of operations resulted in minimized latency because instead of a single node acting as a \textit{chief} having to go through operations one by one: first monitoring the \textit{workers}, then detecting the malicious ones and finally selecting the reliable nodes for training, our technique separates the tasks. The separation of the tasks for some nodes to conducts monitoring and others to conduct training ensures that the miners that are in charge of the training will always have, at every iteration, a reliable set of nodes to select from. Those miners are not concerned of monitoring the behaviour of the nodes because other miners were in charge of doing the monitoring. So they won't waist time and can automatically select reliable nodes to continue training the model. Meanwhile, the other miners in charge of monitoring continue iteration after iteration to assess the behavior of the \textit{workers} and build the pool of reliable nodes. This cooperation between the nodes of the blockchain network permits the implementation of a defense mechanism in an efficient way. Otherwise, the defense would have been too computationally costly in terms of latency and bandwidth. 

\begin{figure}[h]
\begin{center}
  \includegraphics[width=\columnwidth]{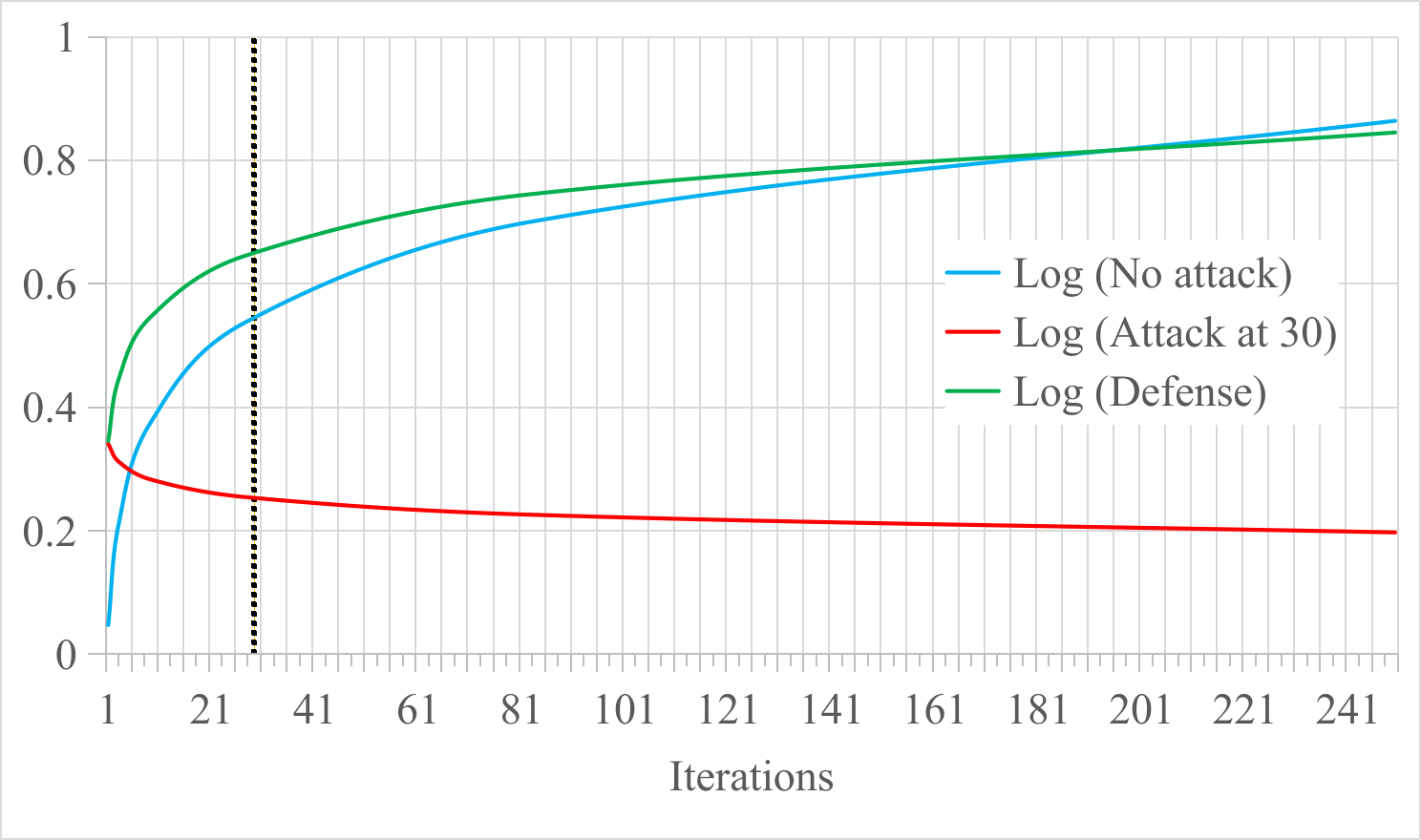}
  \caption{Latency in terms of convergence time of a decentralized FL process for the training of a classifier for transportation mode inference: two attackers.}
  \label{fig:Time_FL_two}
\end{center}
\end{figure}

Furthermore, in order to demonstrate the impact of more than one attacker on the latency of the system, we present in Figure~\ref{fig:Time_FL_two} the convergence time of the system when two attackers are trying to sabotage the training of the model. We notice that our technique performs equally in both scenarios and is agnostic of the number of attackers. On the other hand, as the number of attackers increase, the performance of the miners in the second scenario is slightly more impacted by the poisoning attack and by the number of operations required to complete the process of FL and its cybersecurity. 

\begin{figure}[h]
\begin{center}
  \includegraphics[width=\columnwidth]{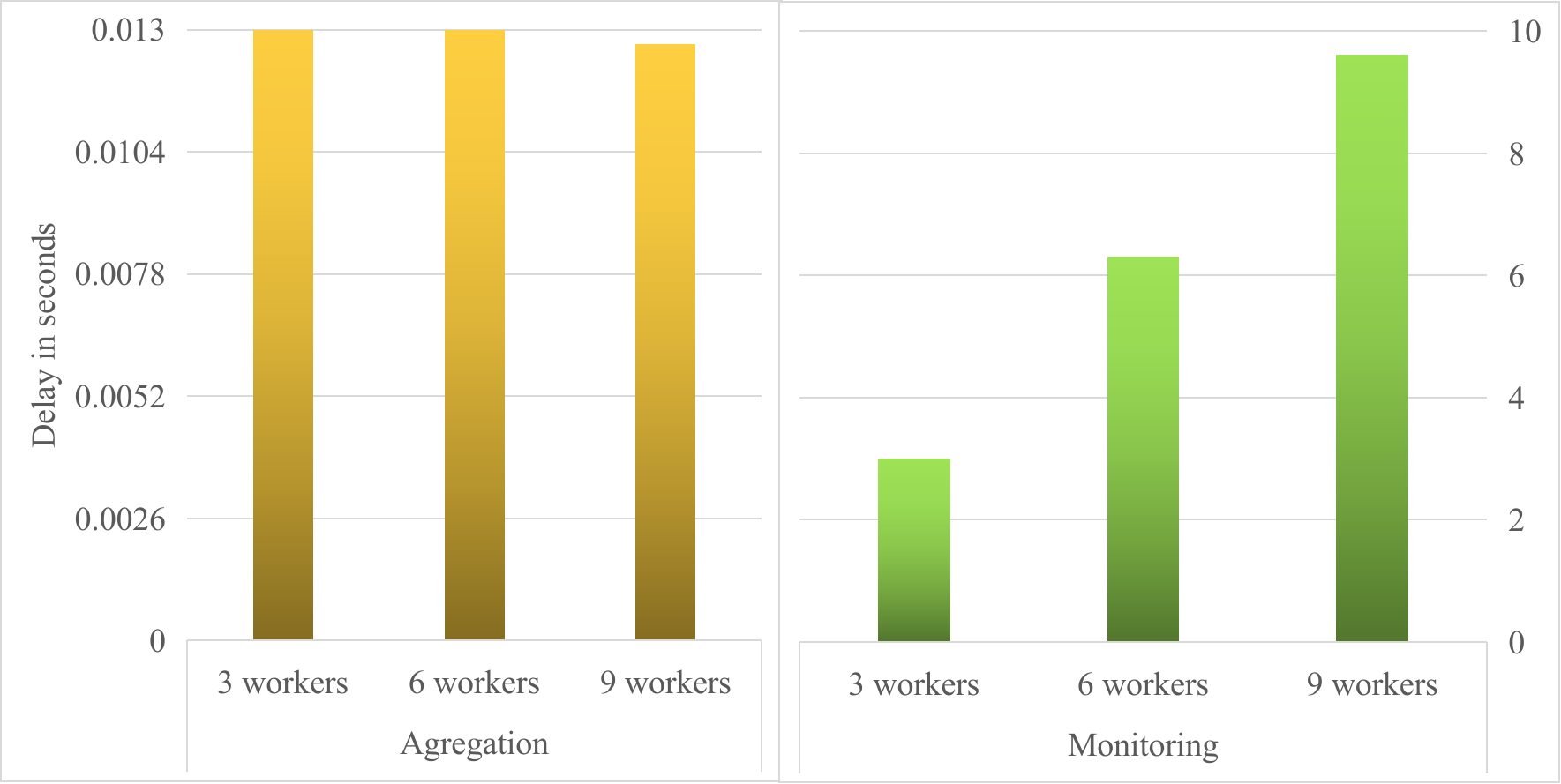}
  \caption{Delay incurred by the operations of the monitoring and detection phases when the network size increases.}
  \label{fig:Miners_MON_FL}
\end{center}
\end{figure}

In terms of scalability, since our technique consists in the decoupling of the monitoring (with \textit{minersMON}) and detection (with \textit{minersFL}) phases, we compare in Figure~\ref{fig:Miners_MON_FL} the time taken by the operations in each phase and that for different network sizes (3, 6 and 9 \textit{workers}). When comparing the overall communication, computation, and consensus delays incurred by the nodes \textit{minersMON} with those of \textit{minersFL}, we notice that even if \textit{minersMON} are impacted by the increase in the number of \textit{workers} in the system, this remains transparent to \textit{minersFL} in the agregation or detection phase since the operations are done in parallel, thus resulting in minimized overall latency.

\section{Conclusion}
\label{sec:conclussions}
We proposed a blockchain-based monitoring technique for poisoning attack detection in decentralized federated learning. Our approach leverages the blockchain network as an immutable security monitoring for the \textit{workers} of the system and is used as a record-keeping privacy preserving pattern collection. The parallelization of the monitoring and detection operations results in minimized latency over the end-to-end communication, computation, and consensus delays incurred during the FL and blockchain operations. Our design achieves robustness and scalability and enables the storage of valuable timely and dynamic local models updates of every \textit{worker} of the system in a transparent and secure manner.  

Our proposed technique can be deployed on resource-constrained nodes such as mobiles or Internet of Things (IoT) devices which have low cost and limited energy. However, for more advanced mobility models, the training may consume significant computation power or bandwidth. To satisfy the resource requirements of such a network, a study of the trade-off between the number of miners of the blockchain network that are attributed to the FL process, the \textit{minersFL} and the \textit{minersMON} that are attributed to the monitoring should be conducted. Moreover, to further ensure that the machine learning model does not suffer from data leakage, meaning from threats such as black-box attacks where malicious
participants can recover arbitrary inputs fed into their devices, in a future work, we are investigating split learning as a countermeasure to achieve more robust privacy preserving model training in a distributed manner. 
\section*{Availability}
Data and code of this study are made publicly available by the authors on~\url{https://github.com/LiTrans/BSMD/tree/master/use_cases/untargeted_poisoning}.
\bibliography{ref_really.bib}
\bibliographystyle{plain}

\end{document}